\documentclass{aa}

\usepackage{amsmath}
\usepackage{amssymb}
\usepackage[usenames, dvipsnames]{color}
\usepackage{footmisc}
\usepackage[modulo,switch]{lineno}
\usepackage{graphicx}
\usepackage{multirow}
\usepackage{natbib}
\usepackage{nicefrac}
\usepackage{pdfpages}
\usepackage[rightcaption]{sidecap}
\usepackage{txfonts}
\usepackage{units}
\usepackage{url}

\usepackage{hyperref}
\hypersetup{
    final=true,
    pageanchor=true,
    colorlinks=true,
    breaklinks=true,
    linkcolor=blue,
    citecolor=blue,
    urlcolor=blue,
    pdfpagemode=UseNone,
    pdftitle={Counter-streaming flows in a giant quiet-Sun filament observed in 
              the extreme ultraviolet},
    pdfauthor={Diercke, Kuckein, Verma, and Denker},
    pdfsubject={Solar Physics},
    pdfkeywords={Method: observational --
             Sun: filaments, prominences --
             Sun: activity --
             Sun: chromosphere --
             Sun: corona --
             Techniques: image processing}}

\newcommand\phn{\phantom{0}}

\bibpunct{(}{)}{;}{a}{}{,} 
\newcommand{\arcdeg}{\mbox{$^{\circ}$}}

\begin{document}

\titlerunning{Counter-streaming flows in a giant quiet-Sun filament}

   \title{Counter-streaming flows in a giant quiet-Sun filament observed in 
the extreme ultraviolet}

   \author{A. Diercke
          \inst{1,2}
          \and
          C. Kuckein \inst{1}
          \and
          M. Verma \inst{1}
          \and
          C. Denker \inst{1}
          }
  \institute{
    Leibniz-Institut f\"ur Astrophysik Potsdam,
    An der Sternwarte 16,
    14482 Potsdam, Germany\\
    \email{adiercke@aip.de}
\and
    Universit\"at Potsdam,
    Institut f\"ur Physik und Astronomie,
    Karl-Liebknecht-Stra\ss{}e 24/25,
    14476 Potsdam,
    Germany 
    }

   \date{Received 31 January 2017; accepted 18 December 2017}

 
  \abstract
{}
   {The giant solar filament was visible on the solar surface from 
2011~November~8\,--\,23. Multiwavelength data from the Solar Dynamics 
Observatory (SDO) were used to examine counter-streaming flows within the spine 
of the filament.}
  {We use data from two SDO instruments,  the Atmospheric Imaging Assembly (AIA) 
and the Helioseismic and Magnetic Imager (HMI), covering the whole filament, 
which stretched over more than half a solar diameter. H$\alpha$ images from the 
Kanzelh\"ohe Solar Observatory (KSO) provide context information of where the 
spine of the filament is defined and the barbs are located. We apply local 
correlation tracking (LCT) to a two-hour time series on 2011~November~16 of the  
AIA images to derive horizontal flow velocities of the filament. To enhance the 
contrast of the AIA images, noise adaptive fuzzy equalization (NAFE) is 
employed, which  allows us to identify and quantify counter-streaming flows in 
the filament. We observe the same cool filament plasma in absorption in both  
H$\alpha$ and EUV images. Hence, the counter-streaming flows are directly 
related to this filament material in the spine. In addition, we use directional 
flow maps to highlight the counter-streaming flows.}
   {We detect counter-streaming flows in the filament, which are visible in the 
time-lapse movies in all four examined AIA wavelength bands ($\lambda$171\,\AA, 
$\lambda$193\,\AA, $\lambda$304\,\AA, and $\lambda$211\,\AA). In the time-lapse 
movies we see that these persistent flows lasted for at least two hours, 
although they became less prominent towards the end of the time series. 
Furthermore, by applying LCT to the images we clearly determine 
counter-streaming flows in time series of $\lambda$171\,\AA\ and 
$\lambda$193\,\AA\ images. In the $\lambda$304\,\AA\ wavelength band, we only 
see minor indications for counter-streaming flows with LCT, while in the 
$\lambda$211\,\AA\ wavelength band the counter-streaming flows are not 
detectable with this method. The diverse morphology of the filament in H$\alpha$ 
and EUV images is caused by different absorption processes, i.e., spectral line 
absorption and absorption by hydrogen and helium continua, respectively. The 
horizontal flows reach mean flow speeds of about 0.5\,km~s$^{-1}$ for all 
wavelength bands. The highest horizontal flow speeds are identified in the 
$\lambda$171\,\AA\ band with flow speeds of up to 2.5\,km~s$^{-1}$. The results 
are averaged over a time series of 90 minutes. Because the LCT sampling window 
has finite width, a spatial degradation cannot be avoided leading to lower 
estimates of the flow velocities as compared to feature tracking or Doppler 
measurements. The counter-streaming flows cover about 15--20\% of the whole area 
of the EUV filament channel and are located in the central part of the spine.} 
{Compared to the ground-based observations,  the absence of seeing effects in 
AIA observations reveal counter-streaming flows in the filament even with a 
moderate image scale of 0\farcs6\,pixel$^{-1}$. Using a contrast enhancement 
technique, these flows can be detected and quantified with LCT in different 
wavelengths. We confirm the omnipresence of counter-streaming flows also in 
giant quiet-Sun filaments.}


   \maketitle

\section{Introduction}

Filaments are clouds of dense plasma that reside at chromospheric or coronal 
heights in the atmosphere of the Sun. Due to differences in temperature between 
the filament and the coronal plasma and absorption of the photospheric 
radiation, filaments are seen as dark structures on the solar disk. Observed 
above the limb, filaments appear in emission as loops and are called prominences 
\citep[e.g., ][]{Martin1998a,Mackay2010}. They are formed above the polarity 
inversion line \citep[PIL,][]{Mackay2010}, which is defined as the border 
between positive and negative polarities of the line-of-sight (LOS) magnetic 
field.

Filaments are usually classified as one of  three types: quiet-Sun, 
intermediate, and active region filaments \citep[e.g., ][]{Engvold1998, 
Engvold2015}. The last are located inside active regions at the sunspot latitude 
belts, whereas  quiet-Sun filaments can exist at any latitude on the Sun 
\citep{Martin1998a,Mackay2010}. Quiet-Sun filaments reach lengths of 
60\,--\,600\,Mm \citep{Tandberg1995}. Some extended quiet-Sun filaments at high 
latitudes ($>$45\arcdeg) are called polar-crown filaments because they 
circumscribe the pole roughly along the same latitude  forming a 
crown~\citep{Cartledge1996a}. They form at the border between the sunspot 
latitude belt and the polar region. The term intermediate filament is used when 
the filament does not fit in the category of quiet-Sun or active region 
filaments.

\begin{SCfigure*}
\includegraphics[width=0.74\textwidth]{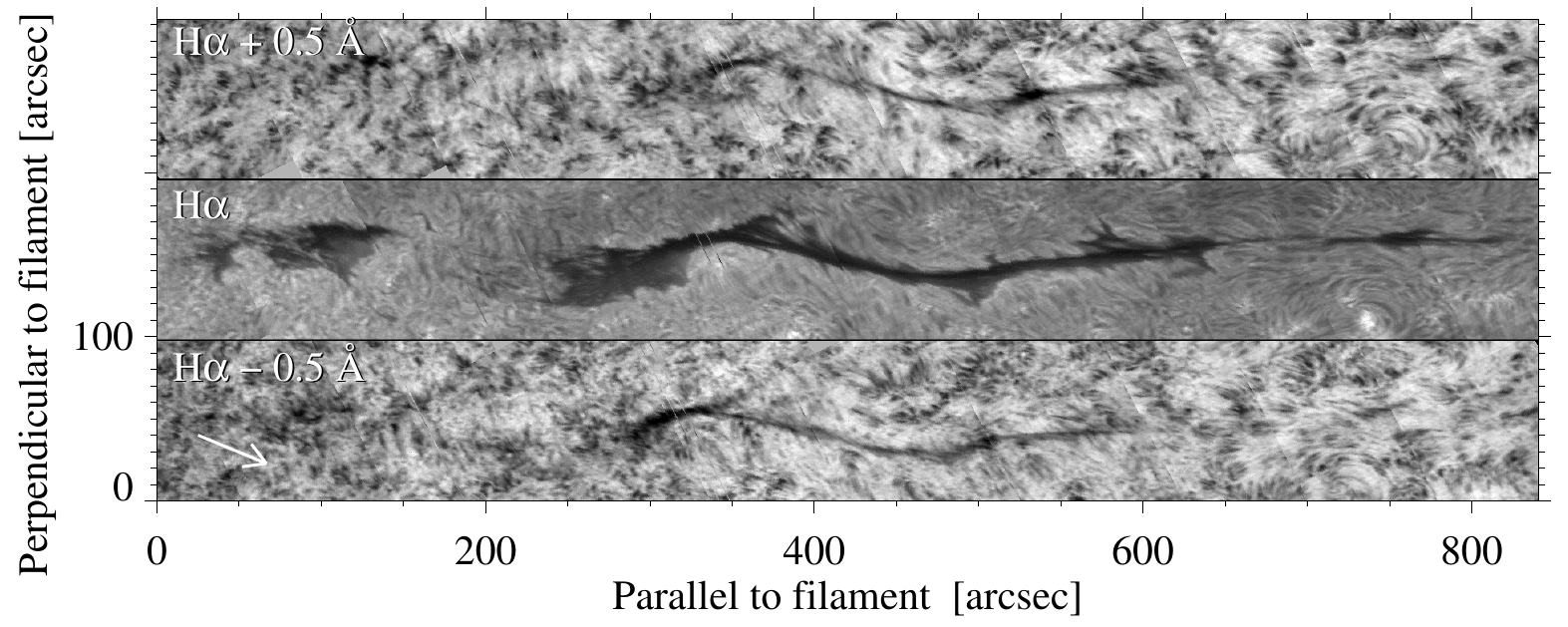}
\caption{Slit-reconstructed image of the VTT observations with the Echelle 
spectrograph at different wavelength positions along the H$\alpha$ line. The 
giant filament extends over about 800\arcsec, displays a gap at the left hand 
side and an EFR appears as a bright area below the thin spine of the filament on 
the right-hand side. The observations were carried out on 2011~November~15 
between 11:38\,UT and 13:20\,UT. From top to bottom: H$\alpha + 0.5$\,\AA, 
H$\alpha$ line core, and  H$\alpha - 0.5$\,\AA. The white arrow points to disk 
center.}
\label{Fig:kuckein}
\end{SCfigure*}

The main part of a filament is the spine. It appears as a dark elongated 
structure. The whole filament is a collection of thin threads, which are aligned 
with the magnetic field lines. Going out sideways from the spine we find  the 
barbs, which are bundles of threads. They are magnetically linked to the 
photosphere \citep{Engvold2001, Lin2005}. Through these barbs, mass can be 
transported into and out of the filament. The barbs and the spine, which contain 
the cool filament plasma, are defined with H$\alpha$ observations. In EUV 
observations the filament has a wider appearance  \citep[e.g.,][]{Aulanier2002, 
Heinzel2003,  Dudik2008}. The filament contains mainly neutral hydrogen and 
helium as well as singly ionized helium. The EUV radiation, emitted by the quiet 
Sun, is absorbed by the Lyman \mbox{H}, \mbox{He\,\textsc{i}}, and 
\mbox{He\,\textsc{ii}} resonance continua. \citet{Anzer2005} state that the 
neutral hydrogen absorbs EUV light below the hydrogen Lyman continuum at 
912\,\AA. In addition, the authors describe that for lower wavelengths, 
absorption through neutral helium (below 504\,\AA) and singly ionized helium 
(below 228\,\AA) occurs. The dark absorption area is defined as the EUV filament 
channel. \citet{Dudik2008} found that the area covered by the EUV filament 
channel matches well with a magnetohydrostatic model of a linear filament flux 
tube. By co-aligning the central part of EUV and H$\alpha$ filaments,  
\citet{Aulanier2002} deduce that the filament channel share some common 
properties in both wavelengths. Both are composed of the same magnetic dips, 
where plasma condensation appear. Furthermore, the authors show that the 
morphology of the filament is the same for observations in several wavelengths 
below 912\,\AA. \citet{Schmieder2008} conclude that the fine structure in 
171\,\AA\ and H$\alpha$ filtergrams is similar because of the comparable optical 
thickness of H$\alpha$ and the \mbox{He\,\textsc{ii}} continuum.

Plasma is moving along the filament. Sometimes these horizontal flows move in 
opposite directions. \citet{Zirker1998} first described this bidirectional 
pattern of flows in a filament as ``counter-streaming'' flows based on H$\alpha$ 
observations. These flows are most clearly seen along threads of the filament's 
spine, but can also occur in barbs \citep{Zirker1998}. Evidently, the formation 
of filaments is a continuous process, where mass is transported into and out of 
the filaments \citep{Martin2001}. Counter-streaming flows are also reported in 
EUV observations, but compared to H$\alpha$ observations they often do not reach 
the combination of high spatial and temporal resolution \citep{Labrosse2010}. 
\citet{Alexander2013} found flows that are oppositely directed in adjacent 
threads in EUV data of the suborbital telescope High Resolution Coronal Imager 
\citep[Hi-C, ][]{Cirtain2013, Kobayashi2014}. They concluded that these flows 
are not visible in EUV images of the Atmospheric Imaging Assembly 
\citep[AIA,][]{Lemen2012} on board the Solar Dynamics Observatory 
\citep[SDO,][]{Pesnell2012} because of the low spatial resolution of AIA. In 
this study, we show that we are able to detect counter-streaming flows in AIA 
images by enhancing the image contrast (Sect.~\ref{sec:results}).

A giant quiet-Sun filament was present on the solar disk in November~2011. The 
main motivation of the present work is to analyze the dynamical aspects of this 
giant filament over a period of two hours. For this purpose we used space 
observations from AIA and the Helioseismic and Magnetic Imager 
\citep[HMI,][]{Schou2012}  on board SDO, and H$\alpha$ observations of the 
Kanzelh\"ohe Solar Observatory (KSO) to define the spine of the filament.  
Furthermore, we demonstrate that it is possible to visualize counter-streaming 
flows in AIA images using an advanced image processing technique, i.e.,  noise 
adaptive fuzzy equalization \citep[NAFE,][]{Druckmueller2013}. The same filament 
was observed with the Echelle spectrograph of the Vacuum Tower Telescope 
\citep[VTT,][]{vonderLuehe1998} on Tenerife, Spain, on 2011~November~15 
(Fig.~\ref{Fig:kuckein}). \citet{Kuckein2016} focused on the spectroscopic 
analysis of the filament. The authors found counter-streaming flows in the barbs 
of the filament, which they related to the mass supply of the filament.

In Sect.~\ref{sec:obs}, we describe the observations and the image processing 
steps for SDO HMI, AIA, and KSO data. In Sects.~\ref{sec:meth} and 
\ref{sec:results}, we present a description of the counter-streaming flows in 
the two-hour time series and use different methods to infer the horizontal 
proper motions along the filament's spine. Furthermore, we focus on the 
visualization and detection of the counter-streaming flows in the spine of the 
filament using local correlation tracking \citep[LCT,][]{November1988}.


\section{Observations}\label{sec:obs}

\begin{figure*}
\includegraphics[width=\textwidth]{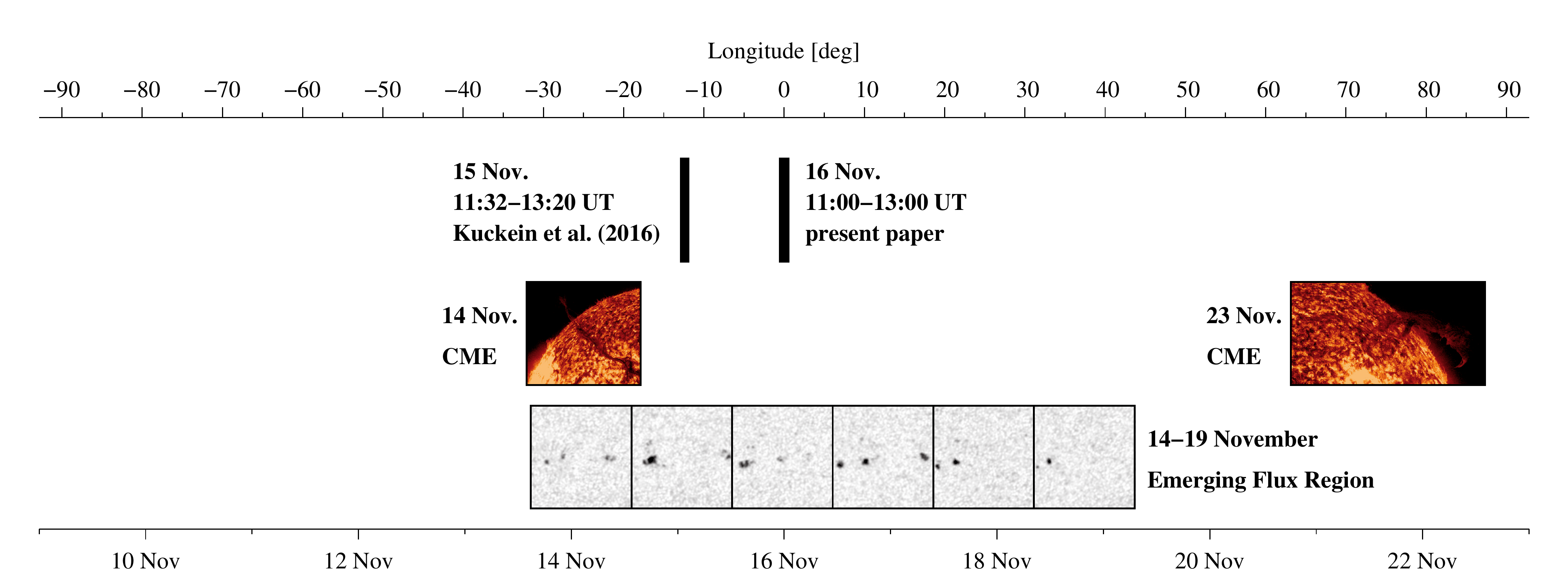}
\caption{ Timeline showing selected events related to the filament  that 
occurred during the filament's passage across the solar disk,  also indicated in 
the lower axis showing the time of observations. The indicated dates refer to 
12:00~UT. The longitudinal position of the middle of the filament is shown at 
the upper axis. The present observations were taken when the filament was 
centered on the central meridian. The  timeline for the CMEs are shown from  
activation to eruption. The EFR, located to the right of the filament, 
is shown at  different evolutionary stages.} 
\label{Fig:timeline}
\end{figure*}

\subsection{Overview}

On 2011~November~8 a giant filament appeared on the solar disk  in the northern 
solar hemisphere. The eastern end of the filament erupted a few days later on 
November~14 as part of a coronal mass ejection (CME).  On October~23 we were 
already able to recognize the filament in $\lambda$304\,\AA\ filtergrams of the 
Solar Terrestrial Relations Observatory \citep[STEREO, ][]{Kaiser2008}. In 
Fig.~\ref{Fig:kuckein} we see the filament in H$\alpha$ as observed with the 
Echelle spectrograph of the VTT on November~15 \citep[][]{Kuckein2014, 
Kuckein2016}. The H$\alpha$ line center image (middle row) shows that the 
filament has a gap on its left side caused by a CME a few hours before. In 
addition, an emerging flux region (EFR) at coordinates (740\arcsec, 10\arcsec) 
is visible as a bright plage in H$\alpha$ connected by some dark arch filaments 
just below the right end of the filament. In HMI continuum images, we see pores 
appearing between November~14\,--\,19 at the location of  the EFR. By 
November~19 both EFR and pores have decayed. Furthermore, we observe many plage 
regions around the filament channel. Much of the filament erupted as a 
non-geoeffective CME on November~22, likely because of a nearby active region 
destabilizing the magnetic field topology. The filament vanished completely one 
day after the CME on November~23. An overview and timeline of the events around 
the filament is given in Fig.~\ref{Fig:timeline}.

The observed giant filament does not fit into the category of polar-crown 
filaments because it covers several tens of degree in latitude and reaches into 
the sunspot latitude belt. Hence, we classify the filament into the wider 
category of quiet-Sun filaments. We examine a phenomenon, which has not been 
reported very often in the past, thus providing us the opportunity to compare 
its properties to regular-sized filaments.
 
\subsection{SDO observations}

On November~16 at around 12:00~UT the filament was centered on the central 
meridian (Fig.~\ref{Fig:timeline}). Hence, we selected this time for further 
analysis. We used AIA and HMI data in order to investigate the spatial evolution 
of the giant filament. The study is based on AIA images taken in the 
EUV-wavelength bands centered at $\lambda$171\,\AA, $\lambda$193\,\AA, 
$\lambda$211\,\AA, and $\lambda$304\,\AA. The  HMI instrument provides 
line-of-sight magnetograms and filtergrams of the continuum intensity. We 
selected observations of two hours from 11:00\,--\,13:00\,UT with a cadence of 
12\,s for the AIA wavelength bands and 45\,s for the HMI magnetograms and 
continuum images.

We downloaded Level 1.0 data, where the data had already gone through the basic 
reduction processes such as flat-field correction and conversion into FITS 
format \citep{Lemen2012}. With the help of a SSW routine, the different plate 
scales and roll angles of all four AIA telescopes were consolidated.  The image 
scale was set to a value of \mbox{0\farcs6\,pixel$^{-1}$} \citep{sdoguide2013} 
for both HMI and AIA images.

After data reduction, the images were prepared for the further analysis.  The 
images have $4096 \times 4096$\,pixels containing the entire solar disk. For the 
purpose of limb-darkening subtraction from the HMI continuum images, we computed 
an average limb-darkening profile. We plotted the intensity values $I$ of each 
pixel as a function of $\mu = \cos\theta$, where $\mu$  is the cosine of the 
heliocentric angle $\theta$, and we fit it with a fourth-order polynomial:

\vspace{-0.45cm}
\begin{equation}
I(\mu) = c_0 + c_1 \cdot \mu + c_2 \cdot \mu^2 + c_3 \cdot
\mu^3 + c_4 \cdot \mu^4 \label{poly1}.
\end{equation}

This profile still includes dark sunspots and bright points in plage areas. To 
remove them from the profile a mask of these regions was created. The mask 
excluded these points, and the procedure was repeated but now with a first-order 
polynomial

\vspace{-0.45cm}
\begin{align}
I(\mu) = c_0 + c_1\cdot \mu.
\end{align}

We expanded the one-dimensional limb-darkening profile to two 
dimensions and subtracted it from the HMI continuum images, which 
leads to a contrast enhancement and a flat background. Making the rough 
assumption that small-scale magnetic fields are perpendicular to the solar 
surface, the HMI magnetograms were divided by $\mu$, which carries out 
a coarse correction of geometric foreshortening to remove projection effects.

\begin{figure}
\includegraphics[width=\columnwidth]{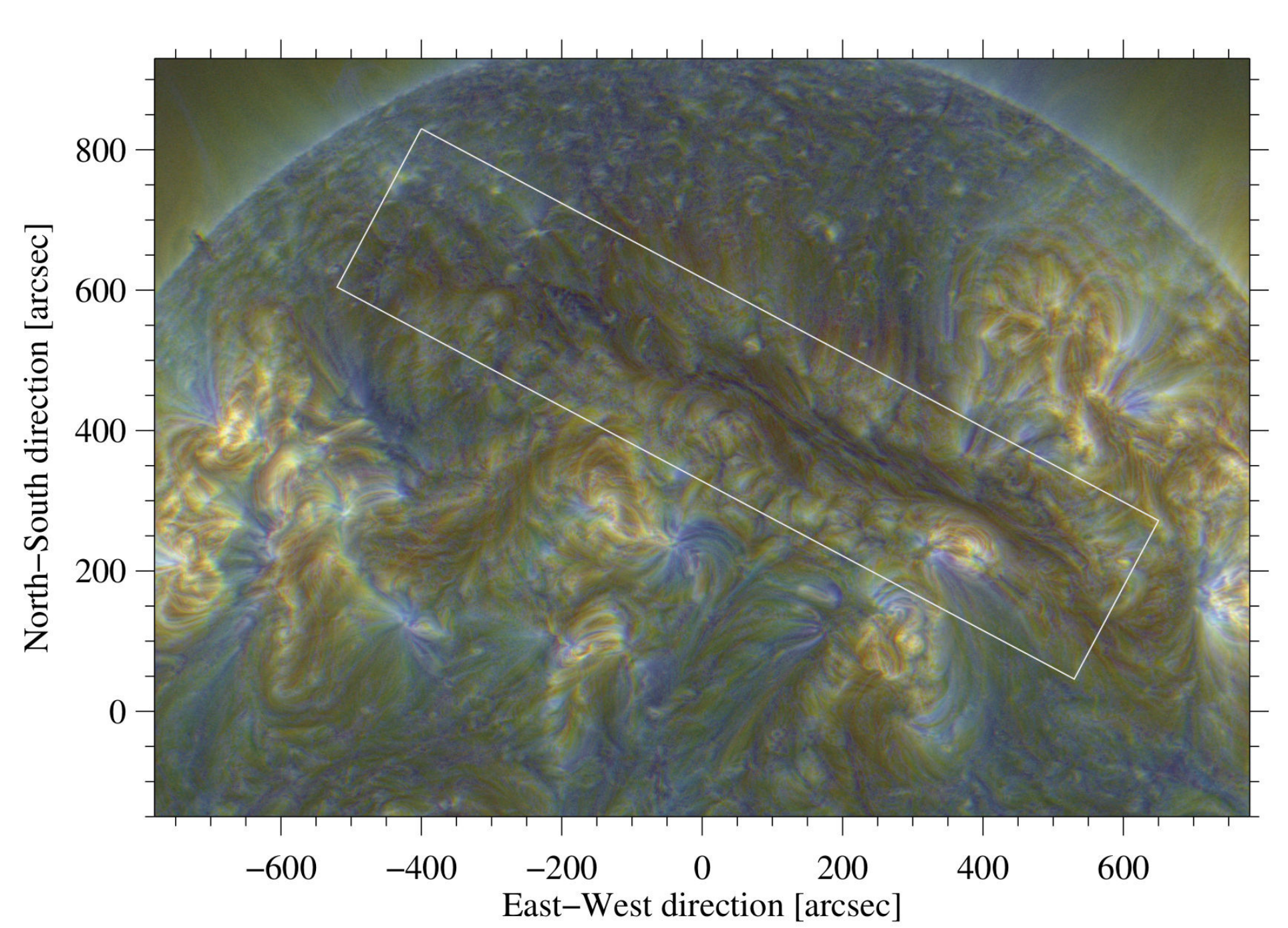}
\caption{False-color image of the giant solar filament at 12:00\,UT on 
November~16. The image is composed of three images obtained in the wavelength 
bands $\lambda$171\,\AA, $\lambda$193\,\AA, and $\lambda$211\,\AA. In the 
composed image different parts of the corona are highlighted. The blue parts 
show lower temperatures in the corona, while the yellow parts show the hotter 
regions. The filament appears as a dark structure enclosed by the white box 
crossing the Sun's central meridian in the northern hemisphere.}
\label{Fig:tricolor}
\end{figure}

Using AIA~$\lambda$304\,\AA\ images, we selected a reference map at exactly the 
time when the center of the filament intersects the Sun's central meridian, 
which occurs  at 12:00\,UT on November~16. To save memory and speed up 
processing we extracted a region of $2600 \times 1800$\,pixels in the northern 
hemisphere (Fig.~\ref{Fig:tricolor}). Routines in the SSW library facilitate a 
straightforward correction of the solar differential rotation. Thus, all data 
were properly aligned.

Because of the spherical shape of the Sun, we implemented geometric 
corrections to obtain a deprojected image of the filament in Cartesian 
coordinates and with square pixels. The origin of this coordinate system 
was  placed at the lower left corner, and the values on the abscissa 
and ordinate are now given in megameters (see Fig.~\ref{Fig:combi}). For this 
purpose, we used a Delaunay triangulation \citep{Lee1980}. Here, 
irregularly gridded data were arranged in triangles by an algorithm. 
The surface values were then interpolated with another algorithm onto 
a regular grid using quadratic polynomial interpolation. This procedure is based 
on \citet{Verma2011}, who used it to correct Hinode  G-band images 
\citep{Kosugi2007}. Furthermore, the filament is rotated  $28$\arcdeg\ 
counterclockwise to have the filament axis horizontal for easier display 
(Fig.~\ref{Fig:combi}). The images are presented with the standard AIA 
color-coding. The images have a size of $3400 \times 1200$\,pixels, where one 
pixel represents an area of $320\,\mathrm{km} \times 320$\,km on the solar 
surface. In Fig.~\ref{Fig:combi}b and c the EUV filament channel has a length of 
around 1000\,Mm.

\subsection{H$\alpha$ observations}

\begin{figure}
\centering
\includegraphics[width=\columnwidth]{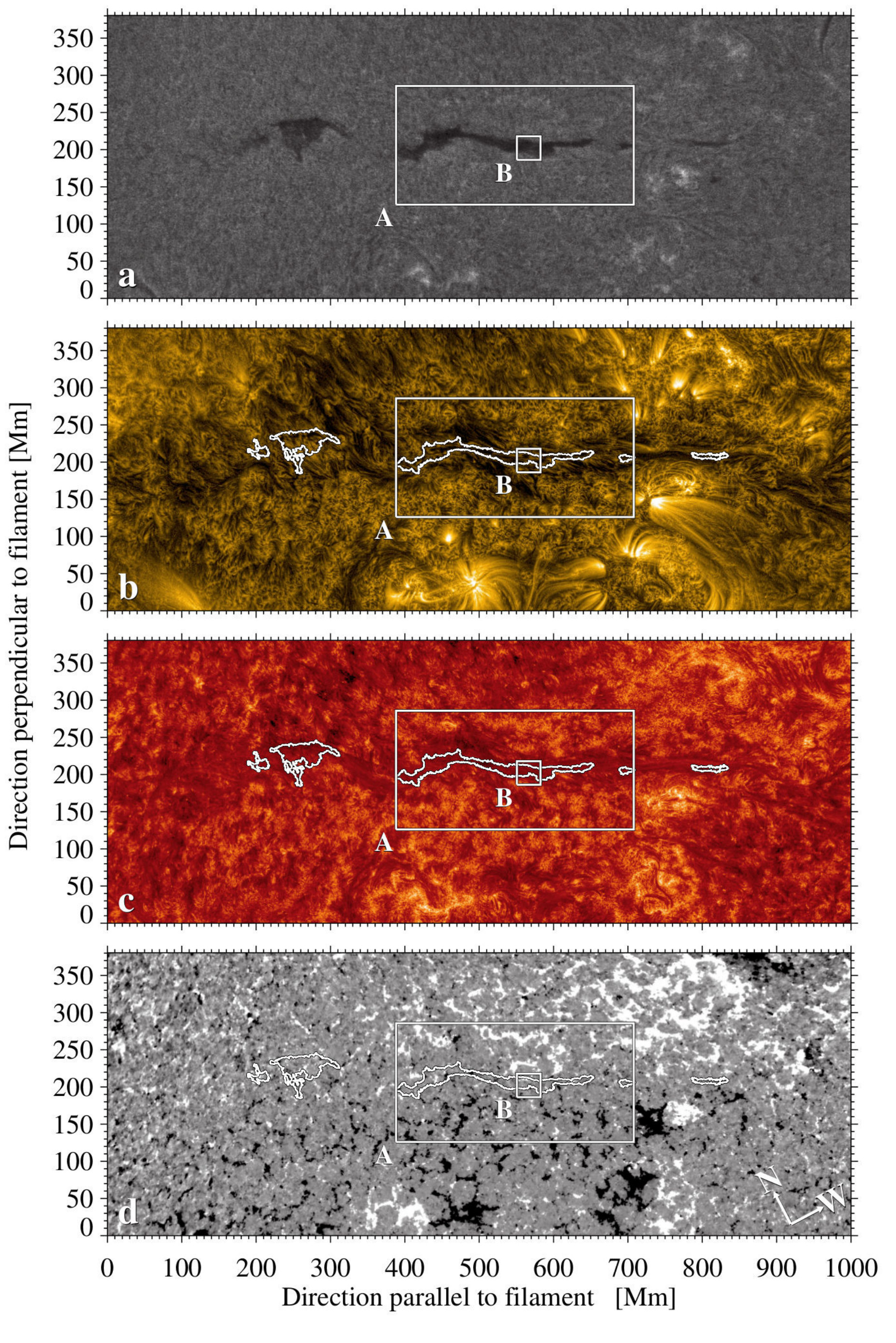}
\caption{The filament at 12:00\,UT on November~16. From top to bottom: (a) 
contrast enhanced H$\alpha$ image from Kanzelh\"ohe Solar Observatory, (b) 
contrast enhanced AIA~$\lambda$171\,\AA\ image, (c) contrast enhanced 
AIA~$\lambda$304\,\AA\ image, and (d) two-hour average of 160 magnetograms 
clipped between $\pm 25$\,G. The contour of the H$\alpha$ filament in panel~(a) 
is plotted in panels~(b), (c), and (d). The white boxes labeled  A and B 
mark the regions that we use later to determine the horizontal proper motions 
(region A, see Figs.~\ref{Fig:171_lct} and \ref{Fig:nafe_small}) and the 
counter-streaming flows (region B, see Fig.~\ref{Fig:171cs})}
\label{Fig:combi}
\end{figure}

We compare the EUV images of the filament with H$\alpha$ observations of the 
Kanzelh\"ohe Solar Observatory (KSO) at 12:00~UT on November~16 
(Fig.~\ref{Fig:combi}a). The contrast of the KSO image was enhanced 
with a $\gamma$-correction. The KSO image was aligned with the SDO 
images and was geometrically corrected with the same procedure as for 
the AIA images. We extracted the contour of the H$\alpha$ filament to 
compare the appearance of the filament with the EUV observations. We see that 
the H$\alpha$ filament is much narrower than the EUV filament channel, as was 
reported in previous studies \citep[e.g., ][]{Anzer2005}. In the H$\alpha$ image 
we recognize the spine of the filament and the barbs. Furthermore, we see the 
gap that was reported in the observations of \citet{Kuckein2016} for the 
previous day (Figs.~\ref{Fig:kuckein} and \ref{Fig:timeline}). In the 
southeastern part of the filament we see further gaps in the spine. The EUV 
filament channel instead appears as a coherent structure where neither the gap 
nor the barbs are visible. The flows, which we describe in the following, are 
directly related to the cold filament material in the spine.

\section{Methods} \label{sec:meth}

To determine the horizontal proper motion along the filament, we use local 
correlation tracking \citep[LCT,][]{November1988} applied to the AIA images. 
Furthermore, we use an advanced image processing tool to enhance the contrast in 
the AIA images, i.e.,  noise adaptive fuzzy equalization 
\citep[NAFE,][]{Druckmueller2013}.

\subsection{Local correlation tracking}

Horizontal proper motions play an important role in the dynamics of solar 
structures. A standard method used to measure horizontal velocity flows is LCT 
\citep[e.g., ][]{November1988, Verma2011}. In this method we determine 
displacement vectors from local cross-correlations between two images of a 
time series. We apply the LCT algorithm of \citet{Verma2011} to the different 
AIA time series to measure the horizontal flow velocities along the spine of the 
filament.

First, we define a Gaussian kernel,

\vspace{-0.45cm}
\begin{align}
g(x',y') = \frac{1}{2 \pi \sigma^2} e^{-\frac{r(x',y')^2}{2 \sigma^2}},
\end{align}
for a high-pass filter, which is applied to suppress gradients related to 
structures larger than granules. The standard 
deviation, 

\vspace{-0.45cm}
\begin{align}
\sigma = \frac{\mathrm{FWHM}}{2\sqrt{2\ln2}}, 
\end{align}
is directly related to the full width at half maximum (FWHM), which has a value 
of 2000\,km (6.5\,pixels). We employ the LCT algorithm to  sub-images $i(x',y')$ 
with a size of $16 \times 16$\,pixels, which refers to a size of 
$5120\,\mathrm{km} \times 5120\,\mathrm{km}$. Thus, the size of the final flow 
map will be reduced by the width of the sampling window. The coordinates $x'$ 
and $y'$ refer to the sub-image, where $r(x',y') = \sqrt{(x'^2 + y'^2)}$. The 
LCT algorithm is used on the whole time period of $\Delta T$ = 120\,min, which 
contains 601 images with a time cadence of $\Delta t$ = 12\,s. In the LCT 
algorithm we use a time cadence of $\Delta t^\prime$ = $8 \times 12$\,s = 
96\,s, in agreement with the results of \citet{Verma2011}.

\subsection{Noise adaptive fuzzy equalization}

The NAFE method is a qualitative method for processing solar EUV images of AIA 
to improve their contrast. In the following the mathematical method is presented 
as it is laid out in \citet{Druckmueller2013}. The NAFE method combines two 
classical methods, unsharp masking and adaptive histogram equalization, but does 
not suffer from their typical problems. By combining the two techniques, the 
resulting images are free of processing artifacts and the noise is well 
controlled and suppressed. The pixels of the original image $A$ depend linear on 
the intensity. 

Applying NAFE to the original image $A$, the resulting image $B$ is then a 
linear combination of the images resulting from the gamma transformation 
$T_{\gamma}(A)$ and the NAFE transformed image $E_{N,\sigma}(A)$:

\vspace{-0.45cm}
\begin{align}
B = (1-w) \ T_{\gamma}(A) + wE_{N,\sigma}. \label{eq:nafe}
\end{align}
The parameter $w$ represents the weight of NAFE. By choosing the value $w = 0$ 
the image passes only through the normal gamma transformation, which is a 
constant 
pixel value transform

\vspace{-0.45cm}
\begin{align}
T_{\gamma}(A) = b_0 + (b_0 + b_1) \left(\frac{a_1 - a_0}{A - 
a_0}\right)^{\gamma}
\label{eq:gamma},
\end{align}
where $a_0$ and $a_1$ are constant input values, representing the zero emission 
intensity and the maximum value encountered in the data. The constants $b_0$ and 
$b_1$ are the minimum and maximum output values. Using lower values for $\gamma$ 
yields darker images, and higher values make the images brighter. In contrast to 
the gamma transform, the image in the NAFE method is passing through a pixel 
value transform, which  varies for each pixel, also depending  on its 
neighborhood \citep[cf. ][]{Druckmueller2013}.

\citet{Druckmueller2013} remarks that the NAFE method shows the best results for 
AIA~$\lambda$171\,\AA\ images. Compared to other methods, this method adds  no 
artifacts to the images and contrast and noise are well controlled. The only 
disadvantage is the long processing time when the method is applied to 
$4\mathrm{k} \times 4$k full-disk images.

The result after applying the NAFE method to the AIA~$\lambda$171\,\AA\ images 
is shown in Fig.~\ref{Fig:combi}b.   Values of $\gamma = 2.6$ and $w = 0.25$ for 
the NAFE parameters provide the best visual results. Evidently, the NAFE image 
shows a higher contrast. The structures in the images are sharper and richer in 
detail. Individual loops are clearly discernible in the EFR and other active 
regions, and single threads are uncovered along the filament's spine.

\begin{figure}
\centering
\includegraphics[width=\columnwidth]{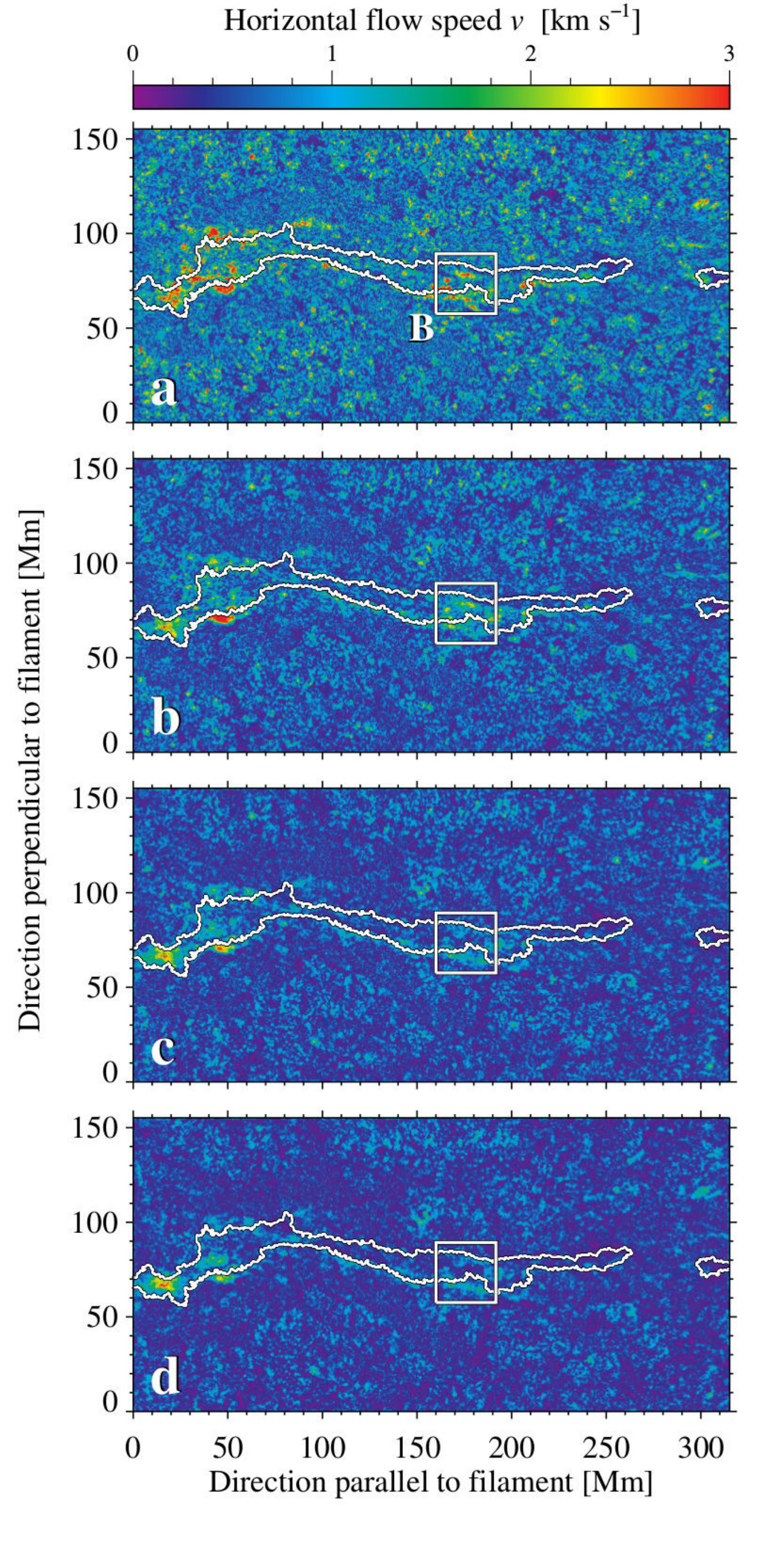}
\caption{LCT flow speed maps of the spine of the filament (region A in 
Fig.~\ref{Fig:combi}). The LCT method was applied to the AIA~$\lambda$171\,\AA\ 
time series for different time intervals $\Delta T$. The time intervals are (a) 
$\Delta T$ = 30\,min, (b) $\Delta T$ = 60\,min, (c) $\Delta T$ = 90\,min, and 
(d) $\Delta T$ = 120\,min. The square-region B shows the $32\,\mathrm{Mm} \times 
32\,\mathrm{Mm}$ region of Fig.~\ref{Fig:171cs}. The white contour outlines the 
filament as seen in H$\alpha$.}
\label{Fig:171_lct}
\end{figure}

We also apply the NAFE method  to the images in the AIA wavelength bands 
$\lambda$193\,\AA, $\lambda$211\,\AA, and $\lambda$304\,\AA\ 
(Figs.~\ref{Fig:tricolor} and \ref{Fig:combi}c, as well as  
Fig.~\ref{Fig:nafe_small} for all four wavelengths). For $\lambda$304\,\AA\ we 
used the  input parameters of $\gamma = 4.4$ and $w = 0.25$. For 
$\lambda$193\,\AA\ and  $\lambda$211\,\AA\ we used the pre-defined values of  
$\gamma$ and $w$ ($\lambda$193\,\AA: $\gamma = 2.6$ and $w = 0.20$; 
$\lambda$211\,\AA: $\gamma = 2.7$ and $w = 0.20$). In summary, for all these 
images we see an increase in the contrast of the images. We identify single 
threads in the spine and around the filament, and in  the loop structure in 
the surrounding plasma.


\section{Results} \label{sec:results}

\subsection{Horizontal flow fields in and around the 
filament}\label{Sec:velo}

To track the horizontal flow fields in and around the filament we apply LCT to 
the contrast enhanced AIA~$\lambda$171\,\AA\ images. The LCT map encompassing 
the whole EUV filament channel (Fig.~\ref{Fig:combi}) possesses a mean velocity 
of $\bar{v} = 0.42$\,km\,s$^{-1}$ with a standard deviation of $\sigma_v = 
0.25$\,km\,s$^{-1}$, and a maximum velocity of $v_{\mathrm{max}}$ = 
3.35\,km\,s$^{-1}$. To inspect the LCT results more closely, we choose a ROI of 
$320\,\mathrm{Mm}\,\times\,160\,\mathrm{Mm}$ within the AIA~$\lambda$171\,\AA\ 
images along the spine of the filament (region A in Fig.~\ref{Fig:combi}). 
First, we study how the time interval $\Delta T$ impacts the velocity. Hence, we 
run the LCT algorithm for the first 15\,min, 30\,min, 60\,min, and 90\,min of 
the time series, and for the whole time series of $\Delta T$ = 120\,min.

In Fig.~\ref{Fig:171_lct}, we display the results for the time intervals $\Delta 
T$ = 30, 60, 90, and 120\,min for LCT applied to region A in 
Fig.~\ref{Fig:combi}. In Table~\ref{tab:lct_171comparison}, the maximum and mean 
values as well as the standard deviation $\sigma_v$ for the velocities are 
listed. For the shorter time intervals the velocities are higher. The shorter 
the time interval, the closer is the flow speed related to the evolution of 
individual features. The highest velocities belong to the 15-minute time 
interval with a maximum velocity of $v_{\mathrm{max}} = 7.7$\,km\,s$^{-1}$. The 
smallest values for the maximum velocity $v_{\mathrm{max}} = 3.3$\,km\,s$^{-1}$ 
are found for the 90-minute time interval. The 120-minute time interval has a 
slightly higher maximum velocity of $v_{\mathrm{max}} = 3.4$\,km\,s$^{-1}$. In 
the 30-minute time interval small groups with high velocities are present in and 
around the spine (Fig.~\ref{Fig:171_lct}a). The longer the time interval, the 
smaller the number of regions with high velocities. In the 60- and 90-minute 
time interval, groups with high velocities are only located in the spine. 
However, in the longer time intervals the overall flow along the spine becomes 
visible (Fig.~\ref{Fig:171_lct}c and d) because the random velocity vectors are 
averaged and only persistent flows along the spine remain. The mean velocity 
shows the same tendency as the maximum velocity. It is highest in the shortest 
time interval with $\bar{v} = 1.2$\,km\,s$^{-1}$ and a standard deviation of 
$\sigma_v =  0.8$\,km\,s$^{-1}$. In the time interval of 120\,min, the mean 
velocity is reduced to nearly one-third of the value in the shorter time 
interval, i.e., $\bar{v} = 0.5$\,km\,s$^{-1}$ and  $\sigma_v =  
0.3$\,km\,s$^{-1}$. \citet{Verma2013} presented in their study similar results 
by varying the time interval of LCT, where it was used on simulated granulation 
patterns. The averaged velocities from the 15- to 120-minute time intervals 
decreased by a factor of three \citep[Fig.~3 in ][]{Verma2013}. In our study LCT 
is applied to a small region of the filament's spine in the corona (region A in 
Fig.~\ref{Fig:combi}). The averaged velocities for the different time intervals 
decreased by nearly 70\% when the time intervals increase. We also recognize the 
absence of the extended flow features reported by \citet{Verma2013} in the 
120-minute time interval compared to the 15-minute time interval.

\begin{table}[t]
\begin{center}
\caption{Comparison of the different horizontal velocities of the 
AIA~$\lambda$171\,\AA\ flow maps for  different time intervals $\Delta T$ of 
the LCT algorithm for region A of $315\,\mathrm{Mm} \times 155\,\mathrm{Mm}$ 
($985 \times 485$\,pixels). We show the maximum velocities $v_{\mathrm{max}}$, 
as well as the mean velocities $\bar{v}$ with the corresponding standard 
deviation $\sigma_v$.}
\begin{tabular}{cccc}
\hline\hline
$\Delta T$    & $v_{\mathrm{max}}$ &  $\bar{v}$ & 
$\sigma_v$\rule[-2mm]{0mm}{6mm}
\\[0ex]
   [min]     &  [km\,s$^{-1}$] & [km\,s$^{-1}$]    &                           
      [km s$^{-1}$]\rule[-2mm]{0mm}{3mm}\\
\hline
 \phn  15   & 7.73 & 1.22 & 0.76\rule[0mm]{0mm}{4mm}\\
 \phn  30   & 6.19 & 0.85 & 0.53\rule[0mm]{0mm}{4mm}\\
 \phn  60   & 4.55 & 0.61 & 0.38\rule[0mm]{0mm}{4mm}\\
 \phn  90   & 3.27 & 0.51 & 0.31\rule[0mm]{0mm}{4mm}\\
      120   & 3.35 & 0.45 & 0.28\rule[-2mm]{0mm}{6mm}\\
\hline
\end{tabular}
\label{tab:lct_171comparison}
\end{center}
\end{table}

\subsection{Counter-streaming flows}

Counter-streaming flows have not yet been clearly seen in AIA EUV images. 
\citet{Alexander2013} compared Hi-C \citep{Cirtain2013, Kobayashi2014} and AIA 
observations of a filament in the $\lambda$193\,\AA\ wavelength band. While the 
Hi-C data showed oppositely directed flows, even the enhanced AIA data did not 
reveal such flows. The authors attributed this discrepancy to the higher spatial 
resolution of Hi-C ($\sim$0\farcs3) compared to AIA~\citep[$\sim${1\farcs5}, 
][]{Lemen2012}. This motivated us to study counter-streaming flows in our 
filament in more detail.

\begin{figure*}
\centering
\includegraphics[width=\textwidth]{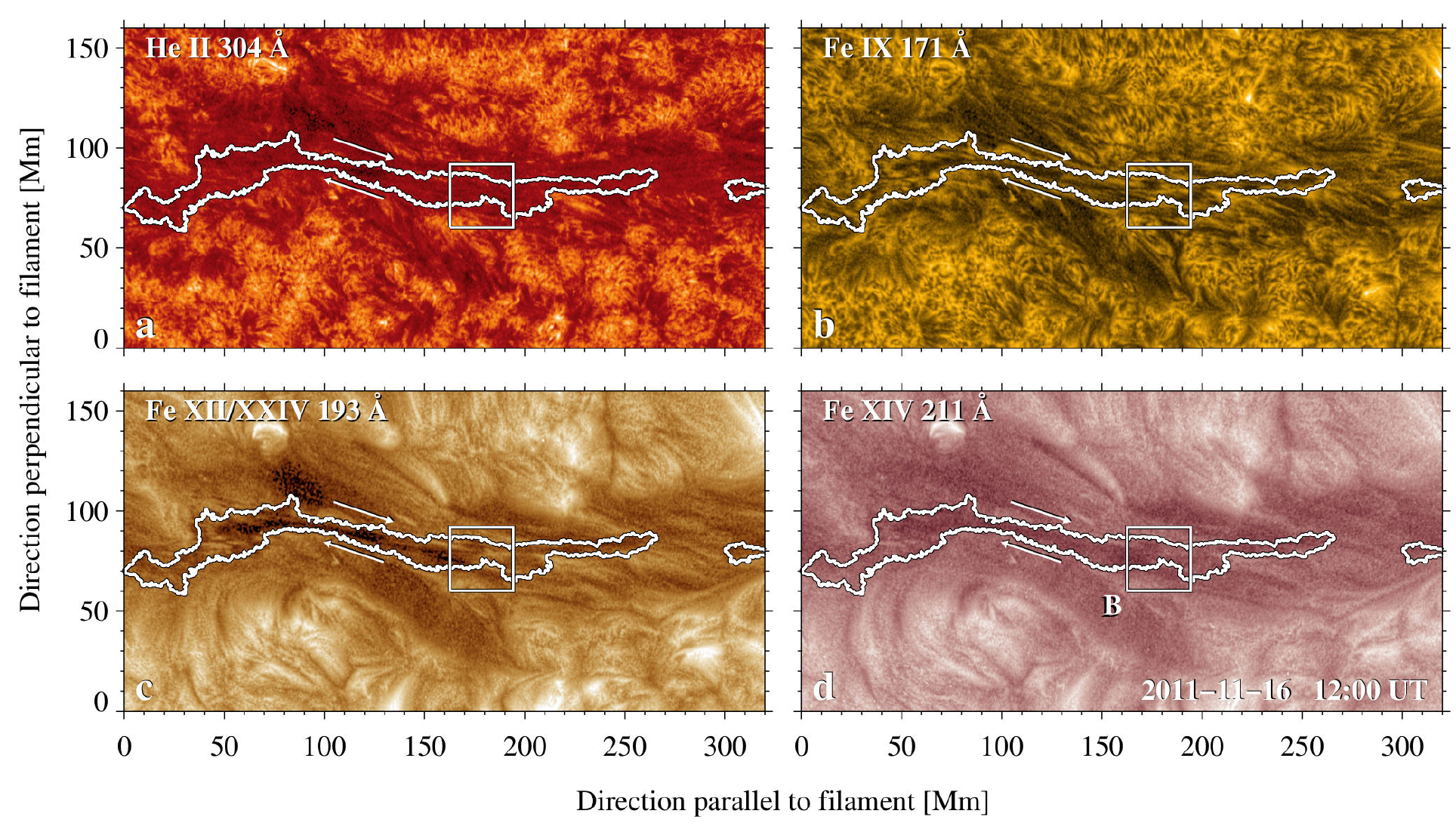}
\caption{NAFE contrast enhanced images (region A with a size of 
$320\,\mathrm{Mm} \times 160\,\mathrm{Mm}$) of different AIA wavelengths at 
12:00\,UT on November~16: (a) $\lambda$304\,\AA, (b) $\lambda$171\,\AA, (c) 
$\lambda$193\,\AA, and (d) $\lambda$211\,\AA. We detect the counter-streaming 
flows in the time-lapse  movies of the high-cadence data set (see online 
material)\protect\footref{note1}. The white square (region B) outlines the 
$32\,\mathrm{Mm} \times 32\,\mathrm{Mm}$ region to which we apply LCT to track 
the counter-streaming flows (Fig.~\ref{Fig:171cs}). The white arrows show 
location and direction of the counter-streaming flows along the spine. The white 
contour encloses the location of the filament's spine in H$\alpha$.}
\label{Fig:nafe_small}
\end{figure*}

We create time-lapse movies for the two-hour time series of the high-cadence 
data set based on the original images and the contrast-enhanced images in the 
four AIA wavelength bands $\lambda$304\,\AA, $\lambda$171\,\AA, 
$\lambda$193\,\AA, and $\lambda$211\,\AA. In all image sequences we detect 
counter-streaming flows along the spine of the giant filament. For the ROI 
(Fig.~\ref{Fig:nafe_small} or region A in Fig.~\ref{Fig:combi}) we provide the 
time-lapse movies of the contrast-enhanced images for all four wavelengths as 
online material.\footnote{\label{note1}A movie for better visualization and 
understanding is available in electronic form at 
\href{http://www.aanda.org/}{http://www.aanda.org}.} The counter-streaming flows 
are present in all four wavelength bands in the same location of the spine. 
Especially at the beginning of the movie, we see a fast south-streaming flow in 
the upper half of the filament's spine directed to the right and parallel to the 
filament. Parallel to it, slower flows in the same direction are observed 
(arrows in Fig.~\ref{Fig:nafe_small}). Simultaneously, we see a north-streaming 
flow in the lower half of the filament's spine directed to the left and also 
parallel to the filament. It appears that these counter-streaming flows are 
faster at the beginning of the time series and slow down towards the end. By the 
end of the time-lapse movie at around 12:45~UT the counter-streaming flows come 
to an end and are no longer clearly visible.

\begin{figure*}
\centering
\includegraphics[width=\textwidth]{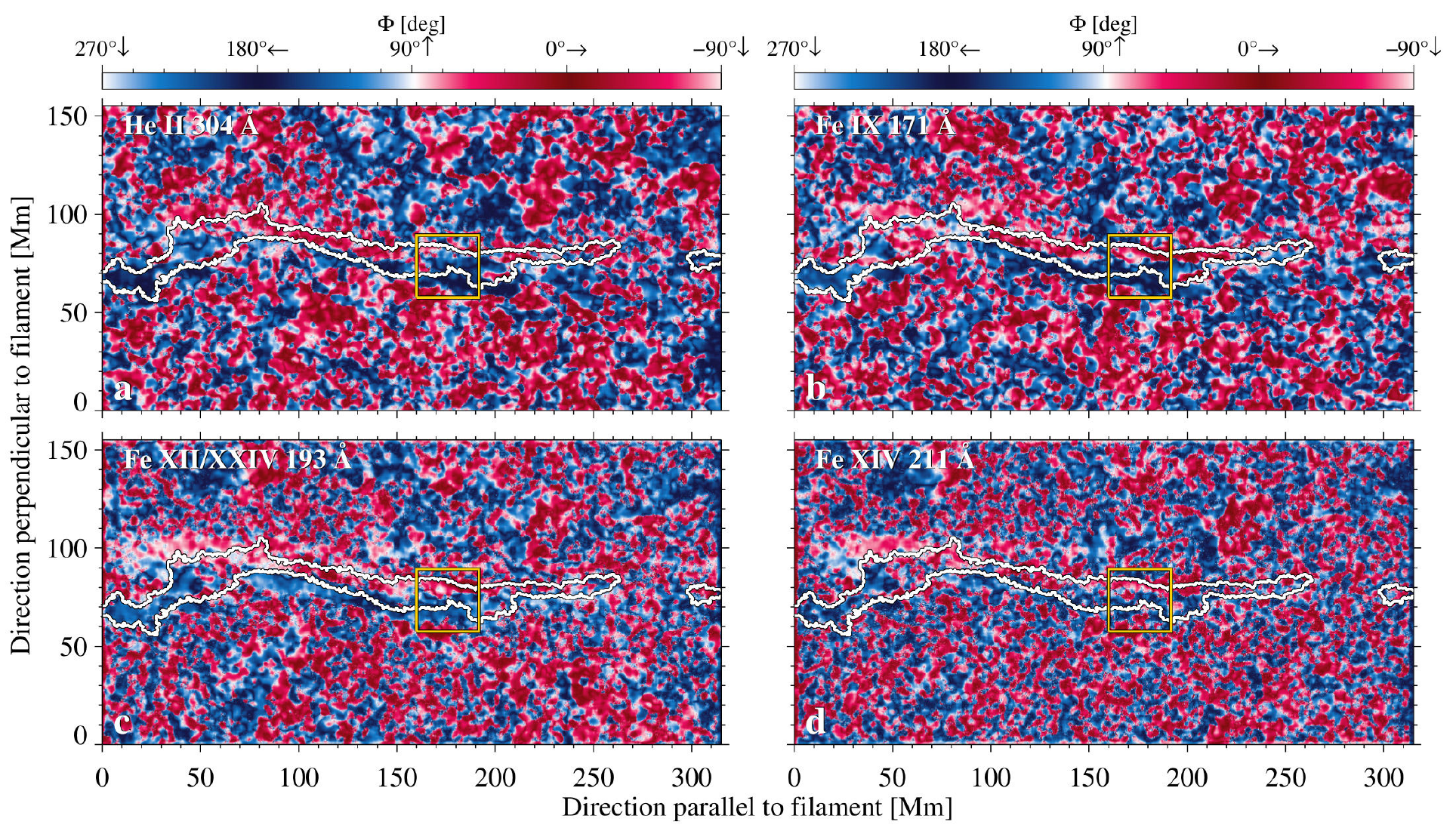}
\caption{Illustration of counter-streaming flows based on the angle 
$\Phi = \mathrm{arccot} \ v_x/v_y$ for the different wavelengths:  (a) 
$\lambda$304\,\AA, (b) $\lambda$171\,\AA, (c) $\lambda$193\,\AA, and (d) 
$\lambda$211\,\AA. Blue (red) indicates that the velocity vector is pointing 
to the left (right). The images were smoothed with a Lee filter.  We 
show the contour of the H$\alpha$ filament in white. The yellow square 
shows a ROI of $32\,\mathrm{Mm}\,\times\,32\,\mathrm{Mm}$, where we 
applied LCT to track the counter-streaming flows (Fig.~\ref{Fig:171cs}).} 
\label{Fig:phi_paper}
\end{figure*}

We use LCT to quantify counter-streaming flows for all four  wavelengths. 
Therefore, we choose a  ROI (Fig.~\ref{Fig:nafe_small}), which has a size of 
$320\,\mathrm{Mm} \times 160\,\mathrm{Mm}$. For LCT we use the first 90\,minutes 
of the two-hour time series because  the counter-streaming flows are best seen 
here. The time-cadence of LCT is $\Delta t^\prime = 96$\,s. With this procedure 
we retrieve the horizontal velocities in the $x$- and $y$-directions, i.e., 
$v_x$ and $v_y$. To illustrate the counter-streaming flows in this region, we 
define the flow direction as the angle $\Phi$ between these two components

\begin{align}
\cot\Phi = \frac{v_x}{v_y} \label{eq:phi},
\end{align}

\noindent i.e., $\Phi = 0\arcdeg$ indicates flows directly to the right and 
$\Phi = 180\arcdeg$ to the left. Flows directly to the top and bottom are given 
by $\Phi = +90\arcdeg$ and $-90\arcdeg$, respectively. The directions of the 
horizontal velocity vectors are color-coded: the darkest red/blue is 
chosen for $\Phi = 0\arcdeg$/$180\arcdeg$. In this way, the direction of flows 
predominantly along the filament's spine are easily perceptible. The results for 
the four wavelengths are shown in Fig.~\ref{Fig:phi_paper}. For the sake of 
clarity, the maps are gently smoothed with a Lee filter \citep{Lee1986} in order 
to emphasize the counter-streaming flows. In addition, we show the contour of 
the filament in H$\alpha$. The overall impression of each map is that the red 
color dominates, i.e., velocity vectors point to the right $\Phi \in 
(-90\arcdeg,\, +90\arcdeg)$. However, in the middle of each map, we see an 
elongated region between coordinates (50\,Mm, 200\,Mm) parallel to the filament 
axis and coordinates (65\,Mm, 85\,Mm) perpendicular to the filament axis, which 
only exhibits velocities directed to the left side of the image (blue). Directly 
above this blue region, we always have a homogeneous red area with velocities 
pointing to the right side. This shows the oppositely directed flows in this 
region of the filament and therefore provides evidence for counter-streaming 
flows. The contour of the filament in  Fig.~\ref{Fig:phi_paper} shows that the 
counter-streaming flows are mainly located in the spine of the filament. This is 
best seen in AIA $\lambda$171\,\AA\ (Fig.~\ref{Fig:phi_paper}b), but also in 
$\lambda$304\,\AA\ (Fig.~\ref{Fig:phi_paper}a) and in $\lambda$193\,\AA\ 
(Fig.~\ref{Fig:phi_paper}c). In  $\lambda$211\,\AA\ (Fig.~\ref{Fig:phi_paper}d) 
the region of counter-streaming flows is barely identifiable in the LCT 
maps. However, the $\lambda$211\,\AA\ movie shows a faint counter-streaming 
flow.

\begin{figure*}
\centering
\includegraphics[width=\textwidth]{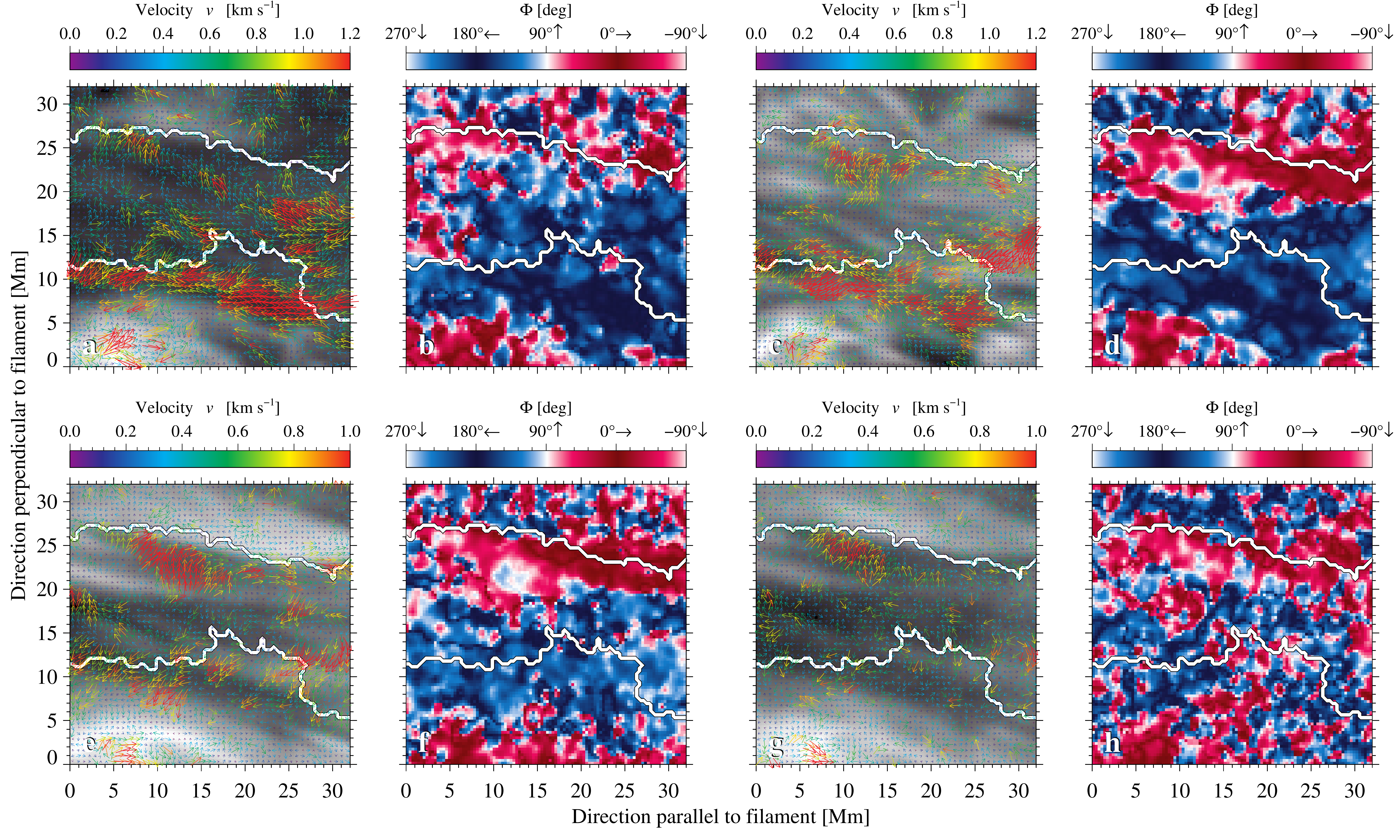}
\caption{Validation of counter-streaming flows: (a, b) $\lambda$304\,\AA, (c, d) 
$\lambda$171\,\AA, (e, f) $\lambda$193\,\AA, and (g, h) $\lambda$211\,\AA. We 
show a region of $32\,\mathrm{Mm} \times 32\,\mathrm{Mm}$ (region B in 
Fig.~\ref{Fig:combi}). The left images (a, c, e, and g) show the summed image of 
each wavelength over the two-hour time series superposed with the horizontal 
velocity vectors derived with LCT. The right images (b, d, f, and h) display a 
map of the angle $\Phi$, which indicates the direction of the velocity vector 
(left is shown in blue and right in red). The white contour outlines 
the filament as seen in H$\alpha$.}
\label{Fig:171cs}
\end{figure*}

To further validate the counter-streaming flows we pick out a small ROI of 
$32\,\mathrm{Mm} \times 32\,\mathrm{Mm}$ at the spine of the filament. This 
region is highlighted as a white and yellow square in Fig.~\ref{Fig:nafe_small} 
and \ref{Fig:phi_paper}, respectively. We show the averaged horizontal velocity 
vectors derived with LCT in Figs.~\ref{Fig:171cs}a, c, e, and g. As the 
background, we show the average image of the two-hour time series. Next to each 
panel, we have the $\Phi$-map indicating the direction of the horizontal 
velocity vectors (Figs.~\ref{Fig:171cs}b, d, f, and h). Figure~\ref{Fig:171cs} 
demonstrates that the counter-streaming flows are best seen in AIA 
$\lambda$171\,\AA\ (Fig.~\ref{Fig:171cs}c). We detect in the lower half of the 
map flows pointing to the left (below 18\,Mm on the $y$-axis). In the upper half 
of the image  flows are directed to the right (above 19\,Mm on the $y$-axis). 
This is clearly evident in the $\Phi$-map (Fig.~\ref{Fig:171cs}d), where the 
lower half of the map is dominated by the color blue, while the upper half is 
dominated by the color red. The same pattern occurs in AIA  $\lambda$193\,\AA\ 
for both the velocity vectors (Fig.~\ref{Fig:171cs}e) and the flow direction map 
(Fig.~\ref{Fig:171cs}f). The wavelength band $\lambda$304\,\AA\ shows a 
different pattern (Fig.~\ref{Fig:171cs}a), where in the lower half of the image  
strong horizontal flows directed to the left still prevail, but in the upper 
half, the flows are directed to the top or only faint horizontal flows  to the 
right are present. The dominance of the horizontal flows directed to the left is 
also visible in the $\Phi$-map (Fig.~\ref{Fig:171cs}b), where the red parts in 
the upper half are not connected with each other. Both panels belonging to AIA 
$\lambda$304\,\AA\ (Figs.~\ref{Fig:171cs}a and b) show no clear evidence for 
counter-streaming flows. Furthermore, we have inspected AIA  $\lambda$211\,\AA\ 
images. In both panels of  Figs.~\ref{Fig:171cs}g and h, we see no clear 
indication of counter-streaming flows. We  glimpse counter-streaming flows in 
the time-lapse movie; however, we cannot verify them with LCT. Most likely, the 
contrast in the $\lambda$211\,\AA\ images is not strong enough to yield a strong 
LCT signal, despite using NAFE on the images to enhance locally the image 
contrast.

\begin{table}
\begin{center}
\caption{Averaged horizontal velocities for region B inside the 
H$\alpha$ contour of the filament's spine separated in the velocities 
predominantly to the left (blue) and right (red) for all four wavelengths. We 
present the maximum horizontal velocities $v_{\mathrm{max}}$, the mean 
horizontal velocities $\bar{v}$, and the corresponding standard deviation 
$\sigma_v$.}
\begin{tabular}{cccccc}
\hline\hline
$\lambda$ & direction & $v_{\mathrm{max}}$ &  $\bar{v}$ & 
$\sigma_v$ & No. of  \rule[-2mm]{0mm}{6mm}
\\[0ex]
[\AA] &  &  [km\,s$^{-1}$] & [km\,s$^{-1}$]    &                           
      [km s$^{-1}$] & pixels \rule[-2mm]{0mm}{3mm}\\
\hline\hline
\multirow{2}{*}{304} & red  & 1.45 & 0.50 & 0.26 & 1268 \rule[0mm]{0mm}{4mm}\\
                     & blue & 2.29 & 0.70 & 0.37 & 3035 \rule[0mm]{0mm}{4mm}\\
\hline
\multirow{2}{*}{171} & red  & 1.60 & 0.67 & 0.31 & 1686  \rule[0mm]{0mm}{4mm}\\
                     & blue & 2.51 & 0.71 & 0.40 & 2617  \rule[0mm]{0mm}{4mm}\\
\hline
\multirow{2}{*}{193} & red  & 1.83 & 0.59 & 0.32 & 1950 \rule[0mm]{0mm}{4mm}\\
                     & blue & 1.47 & 0.56 & 0.28 & 2353 \rule[0mm]{0mm}{4mm}\\
\hline
\multirow{2}{*}{211} & red  & 1.58 & 0.48 & 0.26 & 2087 \rule[0mm]{0mm}{4mm}\\
                     & blue & 1.42 & 0.47 & 0.24 & 2216 \rule[-2mm]{0mm}{6mm}\\
\hline
\end{tabular}
\label{tab:cs}
\end{center}
\end{table}

Applying LCT yields the horizontal velocities for the four EUV wavelengths of 
the small region B (Fig.~\ref{Fig:171cs}). In the following we discuss the 
velocities inside the H$\alpha$ contours of the spine. We categorize the 
velocities into the two main directions: left (blue) and right (red), see 
Table~\ref{tab:cs} for all four wavelengths. For both AIA $\lambda$171\,\AA\ and 
$\lambda$304\,\AA\ the left-directed (blue) flows have higher mean and maximum 
values than the right-directed (red) flows. In addition, for both bands  the 
mean left-directed velocity is virtually identical with 
$\bar{v}_{\mathrm{304\,\AA}} = 0.70$\,km\,s$^{-1}$  and 
$\bar{v}_{\mathrm{171\,\AA}} = 0.71$\,km\,s$^{-1}$. The maximum left-directed 
velocity for $\lambda$171\,\AA\ is higher than the maximum velocity for 
$\lambda$304\,\AA\ with  $v_{\mathrm{max,\,304\,\AA}} = 2.3$\,km\,s$^{-1}$ and 
$v_{\mathrm{max,\,171\,\AA}} = 2.5$\,km\,s$^{-1}$. In comparison, the 
right-directed velocities (red) are lower by about 36\% for both 
$\lambda$304\,\AA\ and $\lambda$171\,\AA. Figures~\ref{Fig:171cs}a and c confirm 
the higher velocities to the left by inspecting the color-code and length of the 
horizontal velocity vectors. The number of pixels used in the averages are 
higher in the blue region with 3035\,pixels in $\lambda$304\,\AA\ and with 
2617\,pixels in $\lambda$171\,\AA. The dominance of the blue part is also seen 
in Figs.~\ref{Fig:171cs}b and d. For $\lambda$193\,\AA, where the 
counter-streaming flows are very prominent, the velocities for the left- and 
right-directed flows are very similar. However, the maximum horizontal velocity 
is somewhat higher in red, in contrast to $\lambda$171\,\AA\ and 
$\lambda$304\,\AA. The mean velocities are nearly the same for both directions 
in $\lambda$193\,\AA\ (Table~\ref{tab:cs}). The number of pixels are also more 
balanced than for the other two wavelengths: 1950\,pixels for red and 
2353\,pixels for blue. In  $\lambda$211\,\AA\ we barely recognize  
counter-streaming flows in the LCT maps (Figs.~\ref{Fig:171cs}g and h). The blue 
and red parts appear balanced, which is also reflected in the number of pixels 
for each direction with 2087\,pixels and 2216\,pixels for the red and blue part, 
respectively. Furthermore, the maximum and mean velocities are very similar, but 
lower than the other wavelength bands (Table~\ref{tab:cs}). The red part has a 
slightly higher maximum horizontal velocity than the blue part, where the high 
velocities appear in a small area between coordinates 
(5\,--\,15\,Mm,\,20\,--\,27\,Mm) in Fig.~\ref{Fig:171cs}g.


\section{Discussions}

The giant filament was visible for more than two weeks and may have even existed 
longer before it rotated onto the visible part of the Sun.  During the 
time of observations of the filament two non-geoeffective CMEs were released. 
The filament lifted  off during the second CME.

We used LCT to  determine the horizontal flow velocities. Applying this 
algorithm to the $\lambda$171\,\AA\ time series of the whole EUV filament 
channel provided a mean horizontal velocity of $\bar{v}_{\mathrm{LCT}} = 
0.42$\,km\,s$^{-1}$ with a standard deviation $\sigma_{\mathrm{LCT}} = 
0.25$\,km\,s$^{-1}$ and a maximum velocity of $v_{\mathrm{max}} = 
3.4$\,km\,s$^{-1}$.  Furthermore, we detected a strong dependency between the 
velocities and the length of the time intervals $\Delta T$ 
(Table~\ref{tab:lct_171comparison}): the longer the time interval, the lower  
the mean velocity value tracked with LCT. In general, the horizontal velocities 
tracked in a filament reach values of 5--30\,km\,s$^{-1}$ \citep{Labrosse2010}. 
Comparing these values to our results, the average velocities we derived with 
LCT are much smaller. By averaging over a time series of two hours, we get the 
global flow field of  persistent flows, and reduce the influence of evolving 
individual features \citep{Verma2013}. Nonetheless, in the shorter time 
intervals of 15\,min or 30\,min we reach maximum velocities of 
6--8\,km\,s$^{-1}$ in the spine of the filament. On the other hand, comparing 
our results with recent LCT studies by \citet{Verma2012} on a decaying sunspot, 
our results are of the same order of magnitude.

\citet{Kuckein2016}  observed the same giant quiet-Sun filament in the 
chromospheric H$\alpha$ line with the VTT on Tenerife. The authors concentrated 
on the analysis of the spectroscopic data. They derived average LOS velocities 
in the spine of $-0.29$\,km\,s$^{-1}$ (upflows). Furthermore, they found 
signatures of counter-streaming flows in the barbs of the filament and proposed 
that the mass supply through the barbs is a good candidate to explain the supply 
with cold plasma in the filament. In the present study, we add to the previous 
results that counter-streaming flows are also present along the spine of this 
giant filament.

We compare the H$\alpha$ filament with observations in EUV. The EUV filament 
channel has a larger extension than the H$\alpha$ filament, as reported 
previously by, e.g., \citet{Dudik2008} and \citet{Aulanier2002}. In the latter 
study, the authors carefully analyzed the relations among filament channels seen 
in various transition-region and coronal lines. They conclude that all EUV 
filament channels observed with wavelengths lower than 912\,\AA\ show the same 
morphology. The authors mention that even for lower wavelengths of 
\mbox{He\,\textsc{ii}}~304\,\AA\ and \mbox{Fe\,\textsc{xii}}~195\,\AA\ their 
interpretation of filament channels is valid. Therefore, it is justified to 
assume that the observed plasma flows in the EUV channels are related to the 
H$\alpha$ filament.

Counter-streaming flows were observed in the visible and   in the UV and EUV 
\citep[for an overview see][]{Labrosse2010}. Most observations of 
counter-streaming flows are obtained in cool-plasma observations in H$\alpha$. 
\citet{Alexander2013} observed counter-streaming flows at subarcsecond scales 
along adjacent filament threads by direct imaging in the $\lambda$193\,\AA\ EUV 
wavelength band with  high-resolution observations of an active region filament 
with Hi-C. The authors compared their results with AIA $\lambda$304~\AA\ images 
and came to the conclusion that counter-streaming flows cannot be tracked in AIA 
images, due to the coarse image scale. Nevertheless, by applying an image 
enhancement technique (NAFE), we were able to clearly track counter-streaming 
flows along the spine of a giant filament with data from the SDO/AIA instrument 
in different wavelength bands. We examined time-lapse movies of the 
contrast-enhanced images in the four AIA wavelength bands $\lambda$171\,\AA, 
$\lambda$193\,\AA, $\lambda$211\,\AA, and $\lambda$304\,\AA\ and detected 
counter-streaming flows in all four wavelength bands along the spine of the 
filament. Even in the unprocessed images, under careful inspection, the 
counter-streaming flows are present, but at the visual detection limit. Hence, 
image enhancement techniques are crucial in order to detect this phenomenon, 
otherwise it might escape detection. Only with the processed images are we  able 
to quantify the counter-streaming flows with LCT. We see differences in the 
morphology of the filament in the four EUV wavelength bands, which possibly can  
be explained by the different absorption characteristics of the EUV wavelength 
bands with respect to the hydrogen and helium continua.

 A more detailed analysis was carried out by tracking counter-streaming flows 
using LCT on a smaller region in the middle of the spine (region~A in 
Fig~\ref{Fig:combi}). We created a map of the dominant flow direction 
(Fig.~\ref{Fig:phi_paper}). With this direction map, we identify regions that 
exhibit clear counter-streaming flows in the wavelength $\lambda$171\,\AA, 
$\lambda$193\,\AA, and $\lambda$304\,\AA. In $\lambda$211\,\AA, 
counter-streaming flows are also seen, but they are much weaker. To validate 
these results, we analyzed in detail an even smaller region (region~B in 
Fig.~\ref{Fig:combi}). The $\lambda$171\,\AA\ and $\lambda$193\,\AA\ filtergrams 
show the most prominent counter-streaming flows. However, for $\lambda$304\,\AA\ 
we can only detect strong flows in one direction. For $\lambda$211\,\AA\ we 
detect a flow only to the right of the filament. We speculate that the reason 
for not detecting any counter-streaming flows in $\lambda$211\,\AA\ is the low 
contrast in the filtergrams, which complicates tracking with LCT.

Counter-streaming flows were first detected by \citet{Zirker1998} in the spine 
of a quiet-Sun filament by examining the blue and red wings of H$\alpha$ 
filtergrams. The opposite directed flows appear in different threads of the 
filament, not only in the spine but also in the barbs of the filament \citep[as 
seen in the barbs of our filament by ][]{Kuckein2016}. In conclusion, 
\citet{Zirker1998} propose that counter-streaming flows are common features in 
all quiet-Sun filaments. Counter-streaming flows are reported not only in 
quiet-Sun filaments, but also in active region filaments as reported by 
\citet{Alexander2013} and now also for extremely large filaments like  the one 
presented in this work. \citet{Lin2011} agrees with the statement that 
counter-streaming flows are common phenomena. \citet{Qiu1999} considers 
counter-streaming flows as a key feature to hold cool plasma stable inside 
filaments at coronal heights. These authors ascribe the counter-streaming 
motions to small-scale pressure imbalances that are located at the footpoints of 
the filament. These imbalances are caused by an unknown mechanism of the 
magnetic field. Inspecting the magnetograms, the overall magnetic configuration 
of the filament is stable. \citet{Lin2003} detect counter-streaming flows in 
H$\alpha$ filtergrams in the wings at $\pm0.25$\,\AA\ and $\pm0.45$\,\AA\ from 
line center along the spine of an polar crown filament and interprets them as  
evidence of a horizontal magnetic field parallel to the main body of the 
filament.

\citet{Chen2014} classified counter-streaming flows in two ways. The first class 
is ubiquitous small-scale counter-streaming flows, which are associated with 
longitudinal oscillations of single threads. The  filament in the present study 
belongs to the second class and contains large-scale counter-streaming flows, 
which are limited to a certain area and are present in different threads. The 
area covers by visual inspection approximately 15--20\% of the area of the 
entire filament. Inside  this area, the counter-streaming flows were visible for 
at least two hours. The flows are persistent in time, but become less prominent 
at the end of the time series.

There are multiple studies reporting on the velocities in counter-streaming 
flows. \citet{Zirker1998} detected flow speeds in H$\alpha$ between 
5--20\,km\,s$^{-1}$. \citet{Lin2003} provided values of the same range with an 
average value of 8\,km\,s$^{-1}$. Other studies have derived similar values of 
5--15\,km\,s$^{-1}$ \citep{Yan2015b, Schmieder2008, Schmieder2010} or  
10--25\,km\,s$^{-1}$ \citep{Lin2008a, Lin2008b, Shen2015}. In EUV observations 
even higher velocities of up to 70\,km\,s$^{-1}$ have been reported 
\citep{Labrosse2010} where the authors point out a trend to higher velocities 
compared to H$\alpha$ observations. Furthermore, in the study of 
\citet{Alexander2013}, who studied oppositely directed flows in EUV, velocities 
of 70--80\,km\,s$^{-1}$ are tracked for an active-region filament along 
individual threads. Studies directly  related to longitudinal oscillations 
showed slightly higher values of 10--40\,km\,s$^{-1}$ \citep{Li2012}. Typically, 
counter-streaming flows were derived with time-slice diagrams \citep[e.g., 
][]{Lin2008b, Schmieder2010} or measuring of Doppler shifts \citep[e.g., 
][]{Lin2003, Schmieder2010}. In the present study, we derived the horizontal 
flow speeds of the counter-streaming flows for the first time with LCT and  
obtained maximum values of 1.4--2.5\,km\,s$^{-1}$, which are much lower than the 
reported values from other studies. The lower values can be explained by the 
chosen method, where the values are averaged over a time range of 90\,min. 
Another parameter affecting LCT velocities is the size of the sampling window. 
If the choice of FWHM is either too small or too large, it will lead to the  
underestimation of the velocities \citep{Verma2013} by mixing or averaging the 
flows in opposite directions. The FWHM we chose is around the size of a typical 
granule, which has been identified by  \citet{Verma2011} and  \citet{Verma2013}  
as an optimal choice for tracking persistent horizontal proper motions in a time 
series.

Tracking  single features manually in time-lapse movies revealed higher flow 
speeds of 20--30\,km\,s$^{-1}$, but with relatively high measurement 
uncertainties of about 5\,km\,s$^{-1}$. However, intensity variations do not 
necessarily imply plasma motions, thus the concept phase velocity more 
appropriately describes this phenomenon. Looking at a rapidly moving feature at 
the beginning of the time series, we  derived a phase velocity higher than  
100\,km\,s$^{-1}$, but  we cannot distinguish in the time-lapse movies whether 
this fast moving structure is directly related to the filament. In any case, the 
derived velocities of single features are comparable with results in the 
previous studies mentioned above.

Furthermore, we applied LCT to time series of KSO full-disk H$\alpha$ images. 
Because of the low spatial resolution and seeing affecting ground-based 
observations, we were not able to detect counter-streaming flows in the 
H$\alpha$ time-lapse movies or when applying LCT to these data.

\section{Conclusions}

We investigated the dynamics of the giant quiet-Sun filament over a time period 
of about two hours, which is a very narrow time range compared to the total 
lifetime of the filament (Fig.~\ref{Fig:timeline}). To determine the proper 
motions of  the cool filament's plasma observed in different EUV wavelengths, we 
applied the LCT to the high-cadence SDO AIA time series. To increase the 
contrast in the AIA images we used the image enhancement technique NAFE, which 
provided detail-rich images with a high dynamic range. Thanks to this method it 
was  possible to uncover counter-streaming flows along the filament and to 
quantify the velocities using the LCT method. By comparing the area with 
counter-streaming in EUV with H$\alpha$ data of KSO, we can conclude that the 
counter-streaming flows appear along the spine of the filament. We found 
counter-streaming flows in different wavelength bands. The counter-streaming 
flows are tracked most clearly in the wavelength bands $\lambda171$\,\AA\ and 
$\lambda193$\,\AA. Furthermore, the counter-streaming flows appear as persistent 
flows covering 15--20\% of the filament's area and are visible for at least two 
hours. They only become less prominent towards the end of the time series.

In order to analyze if counter-streaming flows are present in all quiet-Sun 
filaments, as stated by \citet{Zirker1998} and \citet{Lin2003}, a larger 
statistic sample of quiet-Sun filaments including polar crown filaments could be 
used. Therefore, it is certainly beneficial to inspect the existing H$\alpha$ 
archives from, e.g., the Chromospheric Telescope \citep[ChroTel, 
][]{Kentischer2008, Bethge2011} and the Big Bear Solar Observatory \citep[BBSO, 
][]{Denker1999, Steinegger2000b}. In addition,  similar studies to those 
presented in \citet{Kuckein2016} can be performed with the Echelle spectrograph 
at VTT in combination with Europe's largest solar telescope GREGOR 
\citep{Schmidt2012} to examine the fine structure of filaments in relation to 
the global size of filaments.


\begin{acknowledgements}
The Solar Dynamics Observatory was developed and launched by NASA. The SDO data 
are provided by the Joint Science Operations Center -- Science Data Processing 
at Stanford University.  H$\alpha$ data were provided by the Kanzelh\"ohe 
Observatory, University of Graz, Austria. This research has made use of NASA's 
Astrophysics Data System. SolarSoftWare is a public domain software package for 
analysis of solar data written in the Interactive Data Language by Harris 
Geospatial Solutions. The NAFE software was kindly provided as freeware by Dr.\ 
Miloslav Druckm\"uller. This study is supported by the European Commission's FP7 
Capacities Programme under the Grant Agreement number 312495. A.D. wants to thank 
the Scientific Writing class at AIP for their helpful comments. We thank the 
anonymous referee for valuable comments significantly improving the manuscript. 
\end{acknowledgements}


\bibliographystyle{aa}
\bibliography{aa-jour.bib,Diercke_GiantFilament.bib}

\end{document}